\documentclass[aps,twocolumn,showpacs,superscriptaddress,preprintnumbers,amsmath,amssymb]{revtex4}

\usepackage{graphicx}
\usepackage{dcolumn}
\usepackage{bm}
\usepackage[dvips]{color} 

\begin{document}


\title{Observation of negative absolute resistance in a Josephson junction}

\author{J.~Nagel}
\affiliation{%
Physikalisches Institut -- Experimentalphysik II and Center for
Collective Quantum Phenomena and their Applications, Universit\"{a}t
T\"{u}bingen, Auf der Morgenstelle 14, D-72076 T\"{u}bingen, Germany
}%
\author{D.~Speer}
\affiliation{%
Fakult\"{a}t f\"{u}r Physik,
Universit\"{a}t Bielefeld,
33615 Bielefeld, Germany
}%
\author{T.~Gaber}
\author{A.~Sterck}
\affiliation{%
Physikalisches Institut -- Experimentalphysik II and Center for
Collective Quantum Phenomena and their Applications, Universit\"{a}t
T\"{u}bingen, Auf der Morgenstelle 14, D-72076 T\"{u}bingen, Germany
}%
\author{R.~Eichhorn}
\author{P.~Reimann}
\affiliation{%
Fakult\"{a}t f\"{u}r Physik,
Universit\"{a}t Bielefeld,
33615 Bielefeld, Germany
}%
\author{K.~Ilin}
\author{M.~Siegel}
\affiliation{%
Institut f\"{u}r Mikro- und Nanoelektronische Systeme,
Universit\"{a}t Karlsruhe (TH), \\
Hertzstra{\ss}e 16, D-76187 Karlsruhe, Germany
}%
\author{D.~Koelle}%
\author{R.~Kleiner}%
 \email{kleiner@uni-tuebingen.de}
\affiliation{%
Physikalisches Institut -- Experimentalphysik II and Center for
Collective Quantum Phenomena and their Applications, Universit\"{a}t
T\"{u}bingen, Auf der Morgenstelle 14, D-72076 T\"{u}bingen, Germany
}%

\date{\today}

\begin{abstract}
We experimentally demonstrate the occurrence of
negative absolute resistance (NAR) up to about
$-1\Omega$ in response to 
an externally applied dc current for a shunted Nb-Al/AlO$_x$-Nb
Josephson junction, exposed to a microwave current 
at frequencies in the GHz range.
The realization (or not) of NAR depends crucially on the 
amplitude of the applied microwave current.
Theoretically, the system is described by
means of the resistively and capacitively shunted 
junction model in terms of a moderately damped,
classical Brownian particle dynamics in 
a one-dimensional potential.
We find excellent agreement of the experimental results 
with numerical simulations of the model.
\end{abstract}

\maketitle


When a static force is applied to a system consisting of mobile
particles, these particles usually move in the direction of the
force, i.~e., they show positive mobility, which leads 
to, e.~g., a positive conductance or resistance in electrical 
systems.
Also well known is the fact that such a system can exhibit 
regions of negative differential mobility/resistance
\cite{White84,Griess90,Balakrishnan95,Cecchi96,Slater97,Zia02}.
However, the absolute mobility/resistance usually remains positive.
The opposite response, i.~e., a motion against the static 
force is termed negative absolute mobility or negative absolute
resistance (NAR). 
This is clearly a quite counter intuitive effect
which, at first glance, might seem even to be in conflict with 
Newton's laws and thermodynamic principles \cite{eic05}.
Yet, nonlinear systems being driven far from equilibrium can
indeed exhibit not only a negative differential resistance but 
also a NAR effect.
Unambiguous and convincing experimental observations of
NAR are still quite scarce, involving systems consisting
of electrons in a sample of bulk GaAs \cite{Banys72}, 
electrons in semiconductor heterostructures \cite{Keay95}, 
electrons in low dimensional conductors \cite{zan01},
and charged Brownian particles in structured microfluidic 
devices \cite{Ros05}.
Apart from the low dimensional conductors, the system was 
always driven out of equilibrium
by means of an ac driving force and then its response 
to an externally applied static perturbation was studied.
On the theoretical side, a considerably larger literature 
is available, most notably on different types 
of semiconductors and semiconductor heterostructures 
\cite{eic05}.
In all those cases (except \cite{Ros05}) NAR is based on purely quantum mechanical 
effects which cannot be transferred into the realm of
classical physics.
For classical systems, a first theoretical demonstration
of the effect was provided in the context of a spatially 
periodic and symmetric model system of interacting Brownian 
particles, subjected to multiplicative white noise 
\cite{Reimann99}.
While each of the different ingredients of the model is
quite realistic in itself, their combined realization
in an experimental system seems difficult.
In particular, the main physical mechanism is based on
collective effects of at least three interacting 
particles \cite{vdb02}.
An entirely different mechanism was later on suggested 
theoretically for a realistic, classical model dynamics
of a single Brownian particle in a suitably tailored,
two-dimensional potential landscape in Ref. 
\cite{Eichhorn02} and subsequently realized experimentally
in Refs. \cite{Ros05,Eichhorn07}.
As a first application of NAR, the separation of
different particle species has been realized
in Ref. \cite{reg07}.
While the underlying basic physical mechanism still
requires at least two spatial dimensions,
very recently, NAR has been analyzed and predicted 
theoretically to occur also in the simplest possible 
case of a single Brownian particle dynamics in 
one dimension \cite{Machura07,Speer07}.
More precisely, two basically different physical mechanisms
capable of generating NAR in such systems have been 
unraveled, namely a purely noise induced effect
in Ref. \cite{Machura07} and a transient chaos 
induced effect in Refs. \cite{Speer07}.
In both cases, an experimental realization by 
means of a Josephson junction subjected to suitable 
dc and ac currents has been proposed.
In this Letter we show that a moderately damped 
Josephson junction being driven by microwaves 
indeed shows NAR of the type predicted in
Refs. \cite{Speer07}. A first hint along these lines 
can be found in Fig. 13 of \cite{Pedersen80}, although without 
further explanation or discussion and no direct reference to the resistively 
and capacitively shunted junction model.

To model the Josephson junction we use the resistively and
capacitively shunted junction model \cite{Stewart68,McCumber68}. 
It describes the equation of motion for the difference $\delta$ of 
the phases of the superconducting order parameter in the 
two electrodes
\begin{equation}
I=\frac{\Phi_0}{2\pi}C\ddot{\delta}
 +\frac{\Phi_0}{2\pi R} \dot{\delta}
 +I_{0} \sin{\delta}+I_{N} \ .
\label{1}
\end{equation}
Here, $C$, $R$, and $I_{0}$ denote the junction 
capacitance, resistance and maximum Josephson current, 
respectively, dots indicate time-derivatives, 
$\Phi_0$ is the magnetic flux quantum, 
and $I=I_{dc}+I_{ac}\sin(\omega t)$ is
the total current applied to the junction, 
consisting of a dc and a high frequency ac component.
The first term on the right hand side of Eq.~(\ref{1}) 
describes the displacement
current $C\dot{U}$, where $U$ is the voltage across the junction, 
and has been rewritten in terms of $\dot{\delta}$ using the 
Josephson relation $\dot{\delta}=2\pi U/\Phi_0$.
The second term describes the current through the resistor $R$, the
third term the Josephson current, and the last term the noise current
arising from Nyquist noise in the resistor.
Its spectral power density is assumed to be white with
$S_I(f)=4k_BT/R$, where $T$ is the temperature and $k_B$
Boltzmann's constant. 
The model (\ref{1}) implicitly assumes that 
magnetic fields created by circulating
supercurrents can be neglected 
(short junction limit).
This holds when the lateral junction dimensions are 
below about $4\lambda_J$, where $\lambda_J=(\Phi_0/4\pi
\mu_0j_0\lambda_L)^{1/2}$ is the Josephson length in 
terms of the critical current 
density $j_0$, the London penetration depth 
$\lambda_L$, and the magnetic permeability $\mu_0$.

By integrating Eq.~(\ref{1}) one obtains 
$\delta$ and, by time averaging, the dc voltage 
$V=\Phi_0\langle\dot{\delta}\rangle/2\pi$ 
across the junction. 
This is the main observable
of our present work, which is measured when 
recording $V(I_{dc})$, the 
current voltage characteristics (IVC).
For numerical simulations, (\ref{1}) can be 
rewritten in dimensionless units by
normalizing currents to $I_0$, voltages to $I_0 R$, 
times to $t_c=\Phi_0/(2\pi I_0 R)$, and hence 
frequencies to $f_c=I_0 R/\Phi_0$, yielding
\begin{equation}
i=\beta_c \ddot{\delta} + \dot{\delta} + \sin{\delta} + i_N\  ,
\label{2}
\end{equation}
where $i=i_{dc}+i_{ac}\sin(\tau f/f_c)$ is the 
normalized applied current,
$\tau=t/t_c$ the normalized time,
$\beta_c=(f_c/f_{pl})^2=2\pi I_0 R^2C/\Phi_0$ the Stewart-McCumber
parameter, $f_{pl}=(I_0/(2\pi\Phi_0C))^{1/2}$ the Josephson plasma
frequency, and $i_N$ the normalized noise current with spectral
density $S_i(f/f_c)=4\Gamma$ and noise parameter 
$\Gamma=2\pi k_BT/I_0\Phi_0$.

In a nutshell, the basic ingredients of NAR as
predicted in \cite{Speer07} are as follows.
The unperturbed deterministic dynamics 
(Eq. (\ref{2}) with $i_{dc}=0$ and $i_N=0$)
exhibits two symmetric attractors, carrying
currents of opposite signs (zero crossing Shapiro steps).
When an external perturbation in the form of a static
bias $i_{dc}$ is applied, a subtle interplay of
this bias force and the dissipation leads to a 
destabilization of that attractor, whose
current points into the same direction 
as the applied bias.
Its remnant is a strange repeller, exhibiting 
transient chaos, hence the name 
``transient chaos induced NAR'' coined 
in \cite{Speer07}.
The actual realization of NAR along these lines
requires a careful choice of model parameters 
in (\ref{2}) within the general regime
of frequencies $f$ comparable to $f_{pl}$ and 
values of $\beta_c$ roughly between 1 and 100.
To obtain precise quantitative results,
we have solved Eq.~(\ref{2}) numerically 
for various such parameter values by integrating 
and averaging over typically $5\cdot 10^3$ periods 
of the ac current.

For experiments, which were performed at $T=4.2\,$K, we used circular
Nb-Al/AlO$_x$-Nb Josephson junctions with an area of
$200\,\mu\mathrm{m}^2$, cf. upper left inset in Fig. 1(a).
The junctions were shunted by a AuPd strip with resistance
$R=1.27\,\Omega$ and integrated in a coplanar wave\-guide. 
We denote the critical current $I_c$ as the maximum dc current for which $V=0$.
In general, $I_c$ is a function of $I_{ac}$ and fluctuations.
By measuring $I_c(I_{ac}=0)$ and matching it with simulations we determined  
$I_0=197\,\mu$A, yielding $I_0 R=250\,\mu$V, 
$f_c=121\,$GHz 
and $\Gamma=9\cdot 10^{-4}$.
The Josephson length  is about $40\,\mu$m, i.~e., well above the
$16\,\mu$m diameter of our junctions assuring the short junction
limit.
The design value of the capacitance was $8.24\,$pF, yielding
$f_{pl}=43\,$GHz, 
and $\beta_c = 7.9$. 
The actual value used in the simulations shown below is somewhat
smaller, namely $\beta_c=7.7$, reproducing particularly  
well the hysteretic IVC in the absence of microwaves.
The transport measurements have been performed with a standard four
terminal method, using filtered leads.
Microwaves between 8 and 35\,GHz, with variable output power $P_m$,
were applied through a semirigid cable that was capacitively coupled
to the $50\,\Omega$ coplanar wave\-guide.
The samples were electromagnetically shielded and surrounded by a
cryoperm shield, to reduce static magnetic fields.

\begin{figure}[tb]
\center{\includegraphics[width=8.6cm]{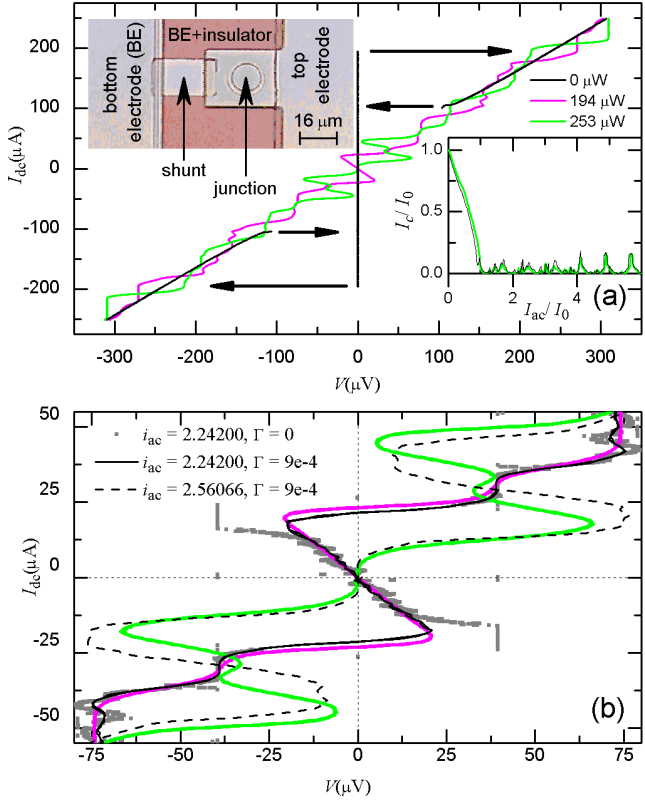}}
\caption{
Current voltage characteristics (IVC) of the Josephson junction
at 4.2\,K in a 19\,GHz microwave field. (a): at 3 levels of applied
microwave power ($0\,\mu$W, $194\,\mu$W and $253\,\mu$W) showing the
effect of negative absolute resistance at $194\,\mu$W
($I_{ac}=435\,\mu$A) and of negative differential resistance 
at $253\,\mu$W ($I_{ac}=497\,\mu$A).
Left inset: image of the Josephson junction.
Right inset: measured (thick green) and
calculated (thin black) dependence of the critical 
current $I_c$ on the microwave current 
amplitude $I_{ac}$.
(b): enlargement of the measured IVCs 
for $194\,\mu$W and $253\,\mu$W, 
together with the simulated IVCs, cf. legend. 
}
\label{fig1}
\end{figure}

Given $I_0$, $R$, $C$, $T$ and $I_{dc}$, all relevant 
model parameters are fixed, with the exception of the
(frequency dependent) coupling factor between the microwave amplitude
$\sqrt{P_m}$ applied from the source and the amplitude $I_{ac}$ of
the ac current induced across the junction.
We have fixed this factor by comparing the measured dependence of 
$I_c(\sqrt{P_m})$ with the calculated curve
$I_c(I_{ac})$, as shown in the right inset of
Fig.~\ref{fig1}, for a microwave frequency of 19\,GHz
($f/f_{c}\approx 0.16$). 
The experimental and theoretical curves are in good agreement.
In particular, the main side maxima can be found, both in experiment
and simulation.
%
%
By adjusting the position of these maxima, we obtain a coupling
factor
$I_{ac}/\sqrt{P_m}(19\,\mathrm{GHz})=1.0\,\mathrm{mA/\sqrt{mW}}$.

Figure \ref{fig1}(a) shows IVCs under $f=19\,$GHz microwave irradiation at
three values of $I_{ac}$.
In the absence of microwaves (black line) the IVC is hysteretic,
exhibiting a critical current of $195\,\mu$A and a return current of
$100\,\mu$A (black arrows).
When the microwave field is applied, the hysteresis decreases with
increasing $I_{ac}$, and step-like features appear on the IVC.
At $P_m=194\,\mu$W ($I_{ac}=435\,\mu$A; magenta line), we observe NAR
with a resistance of $-1.07\,\Omega$, occurring in an interval
$\left|I_{dc}\right|\leq 20\,\mu$A (i.~e., approximately 10\,\% of $I_0$).
When $I_{ac}$ is increased to $P_m=253\,\mu$W ($I_{ac}=497\,\mu$A;
green line) the NAR has disappeared.
However, centered on a voltage which corresponds to the first Shapiro
step ($V_1=\Phi_0 f\approx 39\,\mu$V), regions of negative
differential resistance appear.
In Fig.~\ref{fig1}(b) measured and simulated IVCs for the two
microwave amplitudes $435\,\mu$A and $497\,\mu$A are compared.
For the former case, which is recorded at the microwave amplitude
where the maximum NAR has been observed, the agreement between the
experimental and the theoretical curve is nearly perfect.
For the latter case some small differences can be seen, although the
agreement is still very good.
To demonstrate the origin of the NAR, the grey curve in Fig. 1(b) shows a
simulated IVC for $i_{ac}=2.242\mu A$ and $\Gamma=0$, i.~e., for the noise-free case. 
The curve shows $n=-1$ Shapiro steps to be the cause of NAR, clearly revealing its 
nature to be of the type discussed in \cite{Speer07}. 

\begin{figure}[tb]
\center{\includegraphics[width=8.6cm]{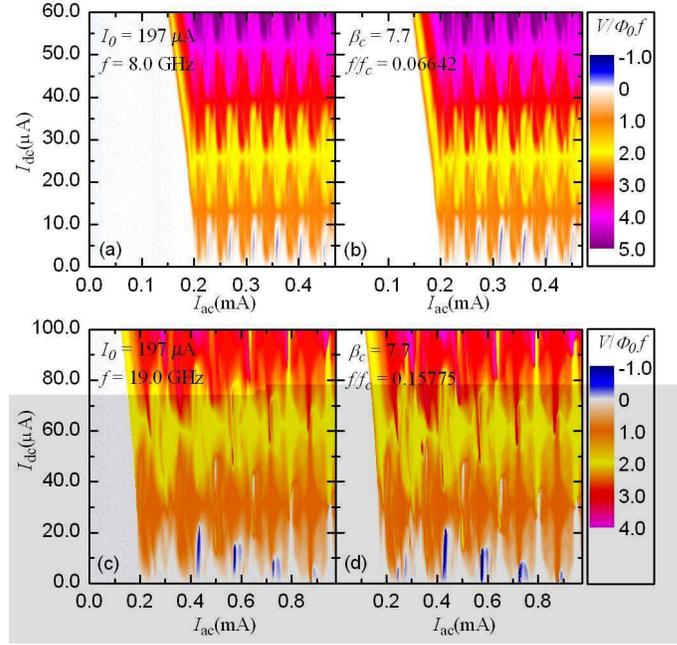}}
\caption{
Contour plot of the normalized dc voltage $V/\Phi_0 f$
across the junction as a
function of dc current $I_{dc}$ and microwave current amplitude
$I_{ac}$.
(a) $f=8\,$GHz, experiment; (b) $f=8\,$GHz, simulation; (c)
$f=19\,$GHz, experiment; (d) $f=19\,$GHz, simulation.
For symmetry reasons, $I_{dc}\mapsto -I_{dc}$ implies
$V\mapsto -V$, hence negative $I_{dc}$ values 
are not shown. 
Blue areas indicate NAR.
}
\label{fig2}
\end{figure}

Figure \ref{fig2} compares in more detail the measured and
calculated dependence of $V$ on $I_{dc}$ and on $I_{ac}$, for two
frequencies (8\,GHz and 19\,GHz).
For $f=8\,$GHz, the comparison between measured  $I_c(\sqrt{P_m})$
and simulated $I_c(I_{ac})$ curves yields a coupling factor
$I_{ac}/\sqrt{P_m}(8\,\mathrm{GHz})=0.33\,\mathrm{mA/\sqrt{mW}}$.
In the graphs, $V$ is normalized to $\Phi_0 f$, yielding an integer
value $n$ for the $n$-th Shapiro step.
Again, the agreement between theory and experiment is very good.
There are at least five $I_{ac}$ intervals where NAR appears at
$f=8\,$GHz, and three such intervals at $f=19\,$GHz.
Within those regimes, the resistance at $I_{dc}=0$  reaches values up
to about $-1\,\Omega$.
In the case of $f=8\,$GHz, the NAR persists up to values of
$|I_{dc}|\approx 10\,\mu$A, for all values of $I_{ac}$ for which NAR
shows up.
In contrast, for $f=19\,$GHz, the $I_{dc}$ interval for NAR decreases
with increasing $I_{ac}$.
When we increased the frequency further to $35\,$GHz, hysteretic
Shapiro steps appeared on the IVC, crossing the voltage axis
($I_{dc}=0$).
As a consequence, NAR ceases to exist both in the experiment 
and the simulations.

\begin{figure}[tb]
\center{\includegraphics[width=8.6cm]{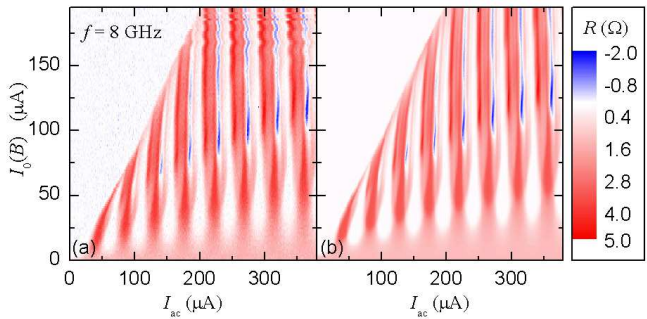}}
\caption{
(a) Contour plot of the experimentally measured
resistance $R:=V(I=5\,\mu\mathrm{A})/5\,\mu$A
as a function of the Josephson current $I_0(B)$ 
and of the microwave amplitude $I_{ac}$. 
$I_0$ has been 
varied by applying a magnetic field 
to the junction.
Graph (b) shows the corresponding simulated plot.
For both graphs parameters at $B=0$ are the same as in Fig.2(a),(b). 
}
\label{fig3}
\end{figure}

In a second series of experiments we applied a magnetic field $B$
parallel to the junction plane in order to tune (decrease) its
Josephson current $I_0$, making it a $B$ dependent function $I_0(B)$ \cite{Barone}.
Thus all $I_0$-dependent parameters entering the normalized equation 
(\ref{2}) aquire a $B$-dependence, in particular $i$, $\beta_c$, $f/f_c$, 
and $\Gamma$. 
Figure \ref{fig3} shows a comparison of the measured and calculated
dependence of the resistance upon $I_{ac}$ and $I_0(B)$.
Again, we find excellent agreement between measurement and theory.
Blue regions indicate NAR.
Their most remarkable feature is that the values of $I_{ac}$, 
for which NAR appears, practically do not depend on $I_0$.
Furthermore, we find that the NAR value can be tuned by $I_0$ 
via an applied magnetic field.
For our junction parameters we find a maximum NAR at 
$I_0\approx (0.4\ldots 0.6) I_0(B=0)$, 
which is increasing with $I_{ac}$.

In conclusion, we have observed negative absolute resistance 
(NAR) of up to about $-1\Omega$ in a shunted Nb-Al/AlO$_x$-Nb
Josephson junction device subjected to microwaves.
To clearly see the effect, a careful choice of parameters 
is required, but still the range of suitable parameters is
quite large.
In all cases, we obtain very good agreement with
theoretical simulations of the resistively and 
capacitively shunted junction model.
Furthermore a closer inspection of the corresponding model
dynamics reveals that the relevant physical mechanism
is of the transient chaos induced NAR type from 
\cite{Speer07}.
The similarity between our Fig. \ref{fig1} 
and Fig. 2 in \cite{Keay95} suggests that with respect
to NAR, purely quantum mechanical band structure
and energy quantization effects may be imitated 
by inertia effects in a purely classical, one 
dimensional noisy dynamics.
Moreover, our Fig. 3 exhibits many features
which are quite similar to the corresponding plots 
in \cite{Speer07}, while the intuitive explanation
of the almost vertical stripe-pattern in Fig. 3
remains as an open problem.
As an application, our present work opens the
intriguing perspective of a new resistor-type
electronic element which is tunable 
between positive and negative resistance
via an easily accessible external control
parameter, e.~g., the amplitude of an
ac driving or an externally applied magnetic 
field in the mT range.\\

This work was supported by the Deutsche Forschungsgemeinschaft
(Grants No.~KO 1303/7-1, RE 1344/5-1, and SFB 613).


\begin{thebibliography}{10}

\bibitem{White84}
S.~R. White and M.~Barma,
J. Phys. A {\bf 17}, 2995 (1984).

\bibitem{Griess90}
G.~A. Griess and P.~Serwer,
Biopolymers {\bf 29}, 1863 (1990).

\bibitem{Balakrishnan95}
V.~Balakrishnan and C.~Van den Broeck,
Physica A {\bf 217}, 1 (1995).

\bibitem{Cecchi96}
G.~A. Cecchi and M.~O. Magnasco,
Phys. Rev. Lett. {\bf 76}, 1968 (1996).

\bibitem{Slater97}
G.~W. Slater, H.~L. Guo, and G.~I. Nixon,
Phys. Rev. Lett. {\bf 78}, 1170 (1997).

\bibitem{Zia02}
R.~K.~P. Zia, E.~L. Praestgaard, and O.~G. Mouritsen,
Am. J. Phys. {\bf 70}, 384 (2002).

\bibitem{eic05}
For a review see 
R. Eichhorn, P. Reimann, B. Cleuren, and C. Van den Broeck, 
Chaos {\bf 15}, 026113 (2005).

\bibitem{Banys72}
T.~J. Banys, I.~V. Parshelyunas, and Y.~K. Pozhela,
Sov. Phys. Semicond. {\bf 5}, 1727 (1972).

\bibitem{Keay95}
B.~J. Keay, S.~Zeuner, Jr. S.~J.~Allen, K.~D. Maranowski, 
A.~C. Gossard, U.~Bhattacharya, and M.~J.~W. Rodwell,
Phys. Rev. Lett. {\bf 75}, 4102 (1995).

\bibitem{zan01}
H. S. J. van der Zant, E. Slot, S. V. Zaitsev-Zotov, and S. N.  Artemenko,
Phys. Rev. Lett. {\bf 87}, 126401 (2001).

\bibitem{Ros05}
A.~Ros, R.~Eichhorn, J.~Regtmeier, T.~T. Duong, P.~Reimann, and D.~Anselmetti,
Nature {\bf 436}, 928 (2005).

\bibitem{Reimann99}
P.~Reimann, R.~Kawai, C. Van den Broeck, and P.~H{\"a}nggi,
Europhys. Lett. {\bf 45}, 545 (1999).

\bibitem{vdb02}
C. Van den Broeck, B. Cleuren, R. Kawai, and M. Kambon,
Int. J. Mod. Phys. C {\bf 13}, 1195 (2002).

\bibitem{Eichhorn02}
R. Eichhorn, P. Reimann, and P. H\"anggi,
Phys. Rev. Lett. {\bf 88}, 190601 (2002);
Phys. Rev. E {\bf 66}, 066132 (2002);
Physica A {\bf 325}.

\bibitem{Eichhorn07}
R.~Eichhorn, A.~Ros, J.~Regtmeier, T.~Tu Duong, P.~Reimann, and D.~Anselmetti,
Eur. Phys. J. Spec. Top. {\bf 143}, 159 (2007);
J. Regtmeier, S. Grauwin, R. Eichhorn, P. Reimann, D. Anselmetti, and R. Ros,
J. Sep. Sci. {\bf 30}, 1461 (2007).

\bibitem{reg07}
J. Regtmeier, R. Eichhorn, T.T. Duong, P. Reimann, D. Anselmetti, and R. Ros,
Eur. Phys. J. E {\bf 22}, 335 (2007).

\bibitem{Machura07}
L.~Machura, M.~Kostur, P.~Talkner, J.~Luczka, and P.~H{\"a}nggi,
Phys. Rev. Lett. {\bf 98}, 040601 (2007).

\bibitem{Speer07}
D.~Speer, R.~Eichhorn, and P.~Reimann,
Europhys. Lett. {\bf 79}, 10005 (2007);
Phys. Rev. E {\bf 76}, 051110 (2007).

\bibitem{Pedersen80}
N. F. Pedersen, O. H. Soerensen, B. Dueholm, J. Mygind,
J. Low. Temp. Phys. {\bf 38}, 1 (1980).

\bibitem{Stewart68}
W.~C. Stewart,
Appl. Phys. Lett {\bf 12}, 277 (1968).

\bibitem{McCumber68}
D.E. McCumber,
J. Appl. Phys. {\bf 39}, 3113 (1968).

\bibitem{Barone}
A. Barone and G. Paterno,
\textit{Physics and Application of the Josephson Effect},
John Wiley and Sons, New York, 1982.

\end{thebibliography}
\end{document}